\documentclass[prl,graphicx, preprint, longbibliograph, superscriptaddress]{revtex4}

\usepackage{graphicx}

\begin{document}

\title{Absorption imaging of a single atom}

\author{Erik W. Streed}
\author{Andreas Jechow}
\author{Benjamin G. Norton}
\author{David Kielpinski}
\email[email:]{d.kielpinski@griffith.edu.au}
\affiliation{Centre for Quantum Dynamics, Griffith University, Brisbane QLD 4111, Australia}
\date{\today}

\begin{abstract}
Absorption imaging has played a key role in the advancement of science from van Leeuwenhoek's discovery of red blood cells to modern observations of dust clouds in stellar nebulas and Bose-Einstein condensates. Here we show the first absorption imaging of a single atom isolated in vacuum. The optical properties of atoms are thoroughly understood, so a single atom is an ideal system for testing the limits of absorption imaging. A single atomic ion was confined in an RF Paul trap and the absorption imaged at near wavelength resolution with a phase Fresnel lens. The observed image contrast of 3.1(3)\% is the maximum theoretically allowed for the imaging resolution of our setup. The absorption of photons by single atoms is of immediate interest for quantum information processing (QIP). Our results also point out new opportunities in imaging of light-sensitive samples both in the optical and x-ray regimes.
\end{abstract}

%%% abstract is 147 words. limit is 150.

\maketitle

\textbf{Introduction}
Absorption of light is a fundamental process in imaging. Since van Leeuwenhoek's discovery of red blood cells, absorption imaging has played a key role in the advancement of science. Modern scientific research, such as astronomical observations of dust clouds in stellar nebula \cite{Padgett-Koerner-HST-young-stars-99} continues to rely upon absorption imaging. The physical limits of absorption imaging thus place an important bound on observable phenomena. In the microscopic limit, recent efforts have measured absorption images from single dye molecules \cite{Gaiduk-Orrit-Single-Mol-Absorption-2010,Celebrano-11}.

An optical image is a map of the effect of an object on the flux of photons from an illumination source. Absorption images originate from the localized depletion of photons by an object. The quality of an absorption image is quantified by the contrast $C = (I_{bkg} - I_d)/I_{bkg}$, where $I_{bkg}$ is the intensity of the background illumination and $I_d$ is the intensity at the darkest point in the image. Slices of an opaque (high contrast) material will become translucent and eventually transparent as the number of absorbing atoms or molecules is reduced towards zero. At the few absorber level the scattering cross-section is generally much smaller than the imaging resolution, resulting in a low image contrast per absorber. In dense objects, from rocks and red blood cells to Bose-Einstein condensates, the large number of absorbers per image pixel more than compensates for the small influence of each absorber. In principle, absorption is the most efficient of all imaging techniques because it measures the total excitation by an absorber. However, for objects of low contrast, the small absorption signal is generally masked by fluctuations in the background illumination. Previous investigations of small quantum systems including fluorescent protein molecules within a living cell \cite{Sako-00}, individual atoms in an optical lattice \cite{Bakr-09}, or trapped-ion crystals \cite{Wineland-87B,Diedrich-87} have relied on imaging a fraction of the fluorescence emission, sacrificing efficiency to reduce background signal and relax resolution requirements.

Quantum mechanics imposes fundamental limits on imaging of a single absorber. For an excitation wavelength $\lambda$ and an optical excitation lifetime $\tau$, no more than a maximum power $P_{\mbox{max}} = hc/( 2\lambda\tau )$ can be absorbed; the object becomes transparent to further illumination. $P_{\mbox{max}}$ is typically on the order of microwatts in the solid \cite{Celebrano-11} or liquid phases and is as small as picowatts for well isolated atoms or molecules  \cite{Tey-08}. The maximum possible scattering cross section at low intensities is $\sigma_0 = \frac{3}{2\pi} \lambda^2$, on the order of the diffraction-limited spot area for imaging at high numerical aperture. A maximum contrast of approximately 92\% can be calculated for this scattering cross section in the low intensity limit \cite{Tey-09}.

A single atom scattering resonant light can closely approach these limits, making it an excellent test system for investigating fundamental limits to imaging. Recent imaging work with single molecules \cite{Gaiduk-Orrit-Single-Mol-Absorption-2010, Celebrano-11} was hampered by the molecules' complicated electronic structure and interaction with a host medium, increasing the theoretical complexity of the system. The maximum optical contrast \cite{Celebrano-11} was limited to 4 ppm and required the use of slow scanning confocal microscopy in conjunction with sophisticated optical signal retrieval techniques. Despite the high total absorption possible with isolated atoms, up to 10\% \cite{Tey-08} in recent spectroscopic measurements, no previous experiment has ever obtained an image of the absorption from a single atom.

 In this work we have imaged the absorption from a single isolated $^{174}Yb^{+}$ atomic ion with an observed maximum contrast of 3.1(3)\%, limited by our imaging resolution and laser cooling dynamics. Our observed contrast is nearly four orders of magnitude better than the previous demonstration of imaging a single absorber \cite{Celebrano-11}. The absorption image contrast and spot size agrees with that expected from a semiclassical model of light scattering from an isolated atom. Hence we realise the maximum theoretically allowed contrast for our imaging resolution. Our approach directly obtains an absorption image in a single exposure, providing a substantial speed advantage over the raster scanning techniques of recent molecular imaging \cite{Gaiduk-Orrit-Single-Mol-Absorption-2010,Celebrano-11}, and is suitable for real-time imaging of dynamic processes, particularly if coupled with pulsed illumination. The absorption of photons by single atoms is of immediate interest for quantum information processing (QIP)  \cite{Specht-Rempe-Cavity-Memory-11,Tey-08, Piro-11}. Absorption imaging of atoms with near-total absorption should be achievable with higher resolution imaging using diffraction-limited optics at higher NA \cite{Tey-09}. Arrays of high-NA optics\cite{Menon-06} are of particular interest for interfacing flying photonic qubits to stationary atomic qubits in the development of massively parallel quantum information processing schemes. Our results also point out new opportunities in imaging of light-sensitive samples both in the optical and x-ray regimes. In particular, the dynamics of chromatin in living cells \cite{Horn-Peterson-Chromatin-Structure-2002} could be imaged without delivering a lethal UV dose.

\textbf{Results}

\textbf{Trapped Ion Imaging System} Our recent advances in high-resolution imaging of trapped ions \cite{Streed-09, Jechow-11} enable us to differentiate the small absorbing area of a single atom from the background illumination. Figure 1 shows the configuration of the experimental apparatus. A $^{174}$Yb$^+$ ion confined in a Paul trap under ultra-high vacuum was laser cooled to a few mK. Light at 369.5 nm was focused to a spot with 4.8 $\mu$m FWHM diameter to simultaneously create a bright illumination field and provide laser cooling. A weak magnetic field along the illumination direction provided a quantisation axis but left the Zeeman levels unresolved. $^{174}$Yb$^+$ has zero nuclear spin and thus no hyperfine structure. After passing the ion, the light was collimated by an in-vacuum phase Fresnel lens (PFL) objective with a numerical aperture of 0.64 and re-imaged onto a cooled CCD camera at up to $615\times$ magnification (see Methods). The highest contrast absorption images were obtained using 4x4 binning of the camera pixels. A spectral interference filter placed in front of the camera eliminated stray light away from the 369.5 nm atomic wavelength. We have previously used this PFL imaging system to perform fluorescence imaging of trapped ions with demonstrated spot sizes as small as 440 nm FWHM diameter \cite{Jechow-11}. Trapped ion lifetimes were several hours, limited by the long term cooling laser mode stability.\\

\begin{figure}[htbp]
\begin{center}
\includegraphics[width=89mm]{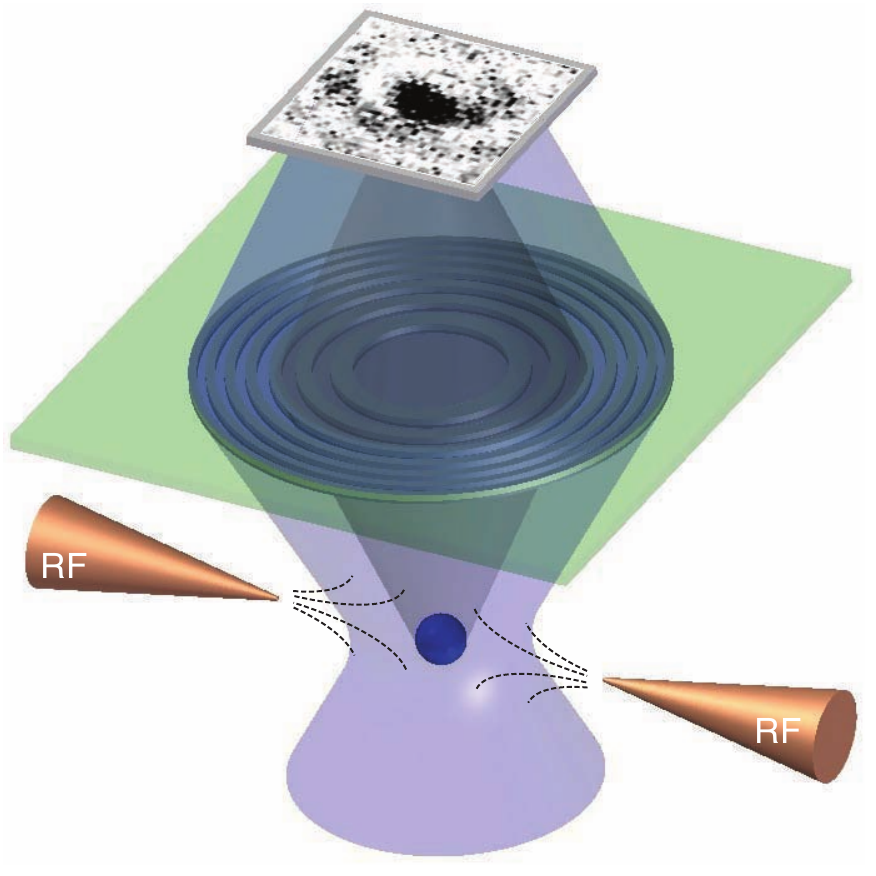}
\caption{Configuration of experimental apparatus. A laser cooled $^{174}$Yb$^+$ ion is confined in a radio frequency Paul trap formed by the electric quadrupole (dashed lines) between two tungsten needles. Resonant illumination at $\lambda$=369.5nm is focused to a spot 4.8 $\mu$m FWHM and absorbed by the ion. The resulting shadow is imaged with a large aperture phase Fresnel objective lens onto a cooled CCD camera at $615\times$ magnification (additional optics omitted for clarity).}
\end{center}
\end{figure}

\textbf{Absorption Images} Figure 2 shows images of a single ion obtained through absorption (Fig. 2B) and fluorescence (Fig. 2D) imaging. Absorption images were obtained by subtracting signal images, for which ion absorption was present, from reference images of the bright field illumination, then normalising each pixel of the subtracted image to its value in the reference image. The reference images were obtained by optically pumping the ion into the meta-stable $D_{3/2}$ atomic state, which does not absorb the 369.5 nm illuminating light. The illumination beam was linearly polarised to eliminate optical pumping effects and its power was stabilised to minimise intensity fluctuations between reference and signal images. Slight modulation of the spatial ion intensity in both images is caused by etalon effects from the interference filter. For quantitative analysis, the absorption images were filtered to reduce background artefacts (see Methods) and then fit to a two dimensional Gaussian. The amplitude of the Gaussian fit gives the contrast, i.e., the probability that a photon passing through the centre of the spot will be absorbed by the ion. A maximum contrast of 3.1(2)\% was observed at low intensity and near optimal laser detuning. The observed absorption spot sizes of 485(69) nm FWHM are consistent with the resolution obtained in fluorescence imaging \cite{Jechow-11}.

Absorption images were acquired with exposure times of 1 s (typical) to 0.05 s (high illumination intensities). Increasing the image magnification increased the number of pixels covered by the absorption spot and allowed for higher intensity exposures within the limited dynamic range of our camera. Absorption image exposure times and signal to noise ratios were limited by technical aspects of beam pointing stability, intensity stabilisation, and dynamic range of the camera. The detailed fluorescence image (Fig. 2D) was obtained by an extended exposure of 60 s and was not limited by beam pointing or intensity stabilisation. Typical fluorescence imaging exposure times in this system are 1 s. Figure 2B has a signal to noise ratio of 4.8(2), typical of the higher contrast absorption images. Undesired deposition of Yb metal from the atomic source onto the imaging viewport reduced the viewport transmission to 20\%. The diffraction efficiency of our phase Fresnel lens is 30\% \cite{Streed-09}. Without these technical limitations acquisition times could be reduced from 1 s down to 60 ms for identical photon fluxes at the camera.

 \begin{figure}[htbp]
 \begin{center}
 \includegraphics[width=178mm]{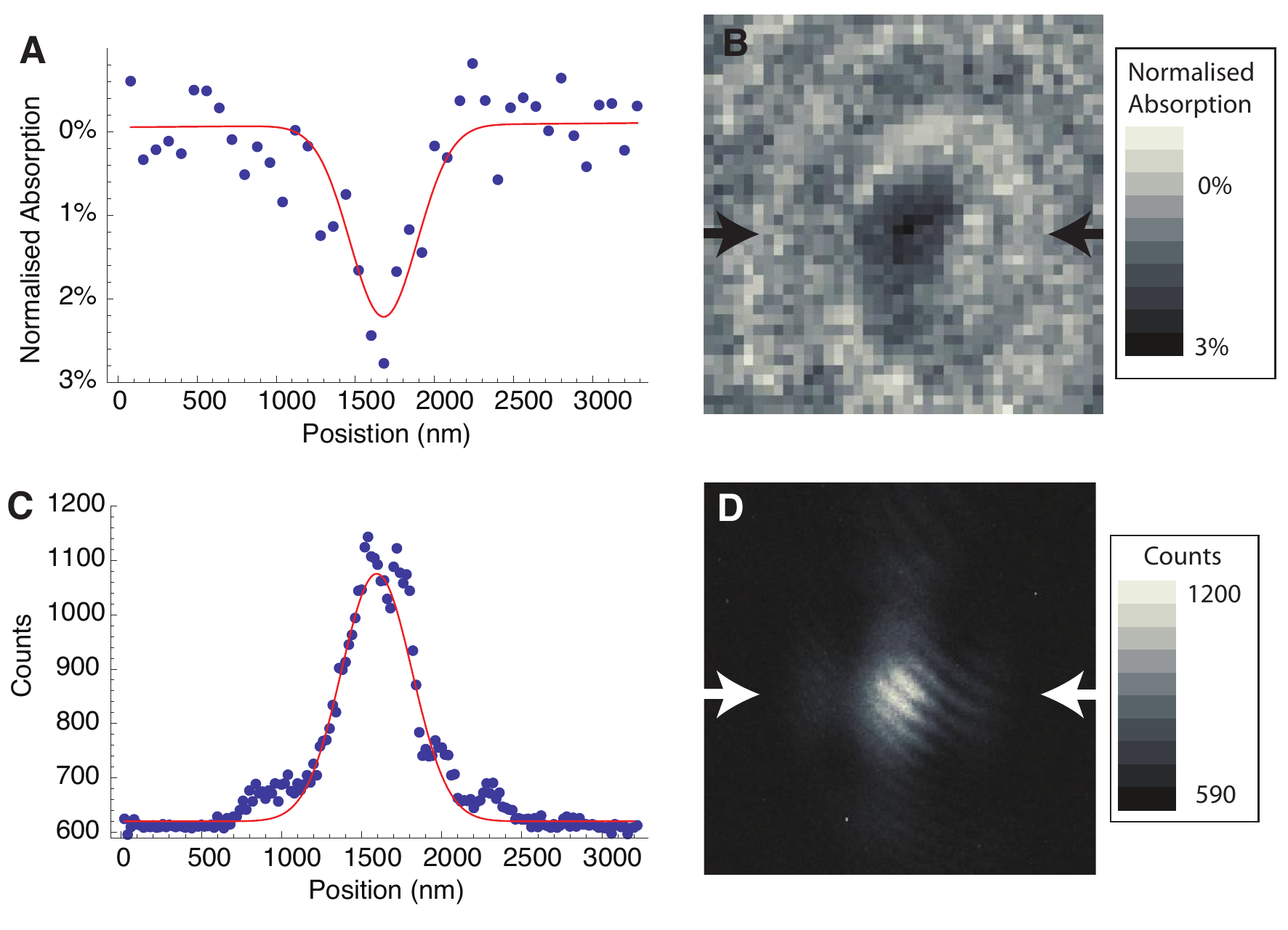}
 \caption{Absorption (A,B) and fluorescence (C,D) images of a single isolated $^{174}$Yb$^+$ ion. Measured at -8 MHz from resonance, near the optimum Doppler cooling detuning for the 369.5 nm transition. A and C are horizontal cross sections taken along the arrows indicated in B and D. Fits are to a two dimensional gaussian over the the whole image. Exposure times are 1 s for B (4x4 pixel binning) and 60 s for D (no binning).}
 \end{center}
 \end{figure}

\textbf{Spectroscopic and Saturation Properties} The dependence of the absorption properties on laser cooling parameters is shown in Figure 3. The frequency dependence of the contrast (Fig. 3A) was measured at  $570 \: \mbox{W} \: \mbox{m}^{-2}$ and follows a Lorentzian response for negative detuning, for which laser cooling is effective. When the detuning is positive, laser heating rapidly increases the ion's motional amplitude and thus the size of the imaged absorption spot, reducing the contrast below detectable levels.  We model this effect by weighting the ideal Lorenztian response with a step function that falls to zero for positive detuning (see Methods). Fitting the data of Fig. 3a yields a Lorentzian linewidth of $35(12) \:\mbox{MHz}$, about twice that expected for an ideal Yb$^+$ ion at rest. Laser cooling dynamics modifies the Lorentzian from that expected for an ion at rest \cite{Norton-11}, shifting the observed detuning of maximum contrast from on resonance to $-8$ MHz. Closer to resonance the increase in scattering rate is accompanied by an increase in the ion spot size from recoil heating, reducing the contrast. These results are consistent with our recent investigations of the dependence of fluorescence image size on laser detuning \cite{Norton-11}. In recent single-atom absorption spectroscopy work \cite{Tey-08} the effects of laser cooling dynamics were minimised by limiting the scattering rate and measurement duration so that only 350 photons were scattered per measurement and only a few measurements could be made before the atom was lost. In contrast, we scatter up to $6\times10^6$ photons per measurement, and due to the exposure times of about 1s our atom could be used for thousands of measurements during the trap lifetime of several hours. Hence our atom is always in equilibrium with respect to the laser cooling dynamics.

Measurement of the total optical power scattered by the atom (described below) as a function of detuning gave an equivalent result with larger uncertainties. The dependence of the absorption on laser intensity (Fig 3B) was measured at -8 MHz detuning, for which we obtain the maximum contrast. The contrast bleaches as the intensity is increased due to saturation of the transition scattering rate. Fitting the saturation data (see Methods) gives a maximum contrast of 3.2(3)\% and a saturation intensity of $585 (128)  \: \mbox{W} \: \mbox{m}^{-2}$, in agreement with the maximum observed contrast of 3.1(3)\% and the theoretically expected saturation intensity of $I_{\mbox{sat}}= \pi h c / (3 \lambda^3 \tau) =508  \: \mbox{W} \: \mbox{m}^{-2}$.

 \begin{figure}[htbp]
 \begin{center}
 \includegraphics[width=80mm]{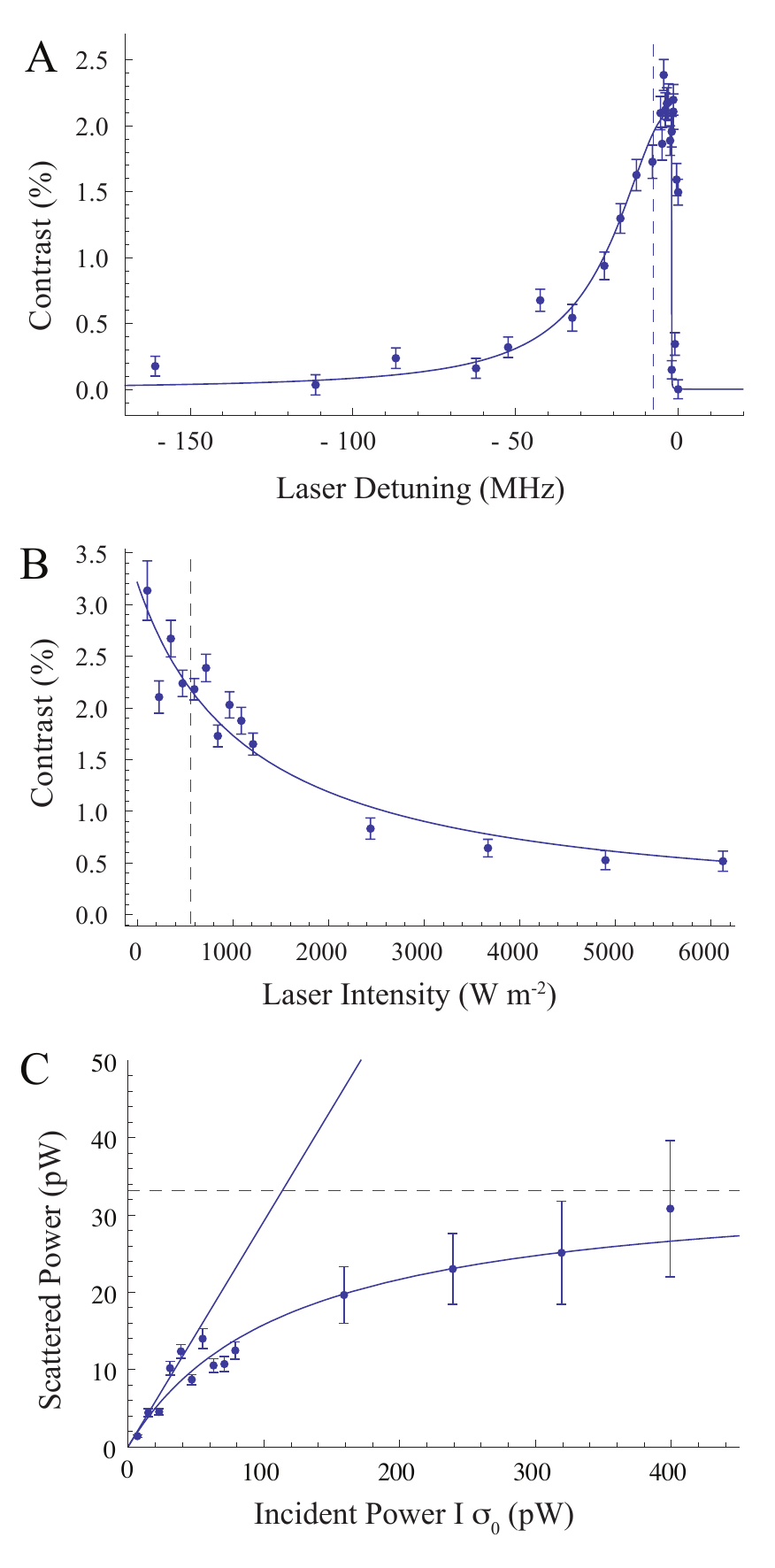}
 \caption{Dependence of absorption properties on laser parameters. The contrast was measured as a function of laser detuning (A) and incident intensity (B) with the other parameter fixed near its optimum value (dashed vertical lines at -8 MHz and $570 \: \mbox{W} \: \mbox{m}^{-2}$ on the alternate graphs). At 0 MHz detuning (rightmost point in (A)) laser cooling ceased and the ion was not detectable until recooled. (C) Total power absorbed by the ion as a function of the incident power on the maximum scattering cross section $\sigma_0$. Dashed horizontal line is the theoretical maximum power absorption $P_{\mbox{max}}= hc/(2\lambda\tau)$. Solid line is the linear extrapolation of the fit at low intensity. Each data point is obtained by fitting to a single image with error bars from the statistical uncertainty in the fit.}
 \end{center}
 \end{figure}

To characterise the ion's power diversion the total optical power scattered by the atom was compared to that incident into the resonant cross-section (Fig 3C). The total optical power scattered by the atom is independent of the imaging resolution and is the product of the contrast, measured absorber spot area, and incident intensity. Fitting to a saturation curve (see Methods) provides an estimate of the maximum absorbed power $P_{\mbox{max}}$ of 34(6) pW, in agreement with the 33 pW expected from the transition wavelength ($\lambda=369.5$ nm) and excited state lifetime ($\tau=8.1$ ns \cite{Olmschenk-09}). In the low intensity limit the ion diverted 30(10)\% of the power incident on the resonant cross section, consistent with the 30\% expected from reductions due to the -8 MHz detuning and differences between the actual transition (J=1/2 electron spin) and an ideal two level system. This difference matches with the difference between our 3\% observed maximum contrast and the 10\% absorption observed in \cite{Tey-08}, which probed a cycling transition (ideal two level system for a particular polarisation) on-resonance at a similar numerical aperture. Since the spatial resolution of our imaging system is larger than the scattering cross section of the atom each pixel includes both scattered and unscattered light, reducing the observed contrast. The agreement between the fitted and theoretical values for Fig 3C indicates that the contrasts measured in Fig 3B achieve their theoretical maximum values for our imaging resolution.\\

\textbf{Discussion}

By achieving absorption imaging with contrast consistent with semiclassical theory in our optical setup, we have demonstrated the maximum available signal extraction per illumination photon. Maximum signal extraction is crucial for imaging light-sensitive samples in the ultraviolet or x-ray regime where exposure to damaging dosages often precludes investigation of phenomena, particularly at short timescales. The phase Fresnel lens elements used in our work are commonly used for X-ray imaging, so that absorption imaging with the maximum theoretically allowed contrast  also appears feasible with X-rays. Cellular processes including gene expression, regulation and transcription depend on the dynamics of nucleic acids as they condense into and out of chromatin strands\cite{Horn-Peterson-Chromatin-Structure-2002}. Timescales for these processes range from 10 ms for spontaneous unwrapping of small DNA nucleosome structures \cite{Li-04} to minutes or longer for mitosis. Nucleic acids have a notable absorption peak in the ultraviolet at 260 nm, giving high absorption contrast with respect to other cell contents. The strong germicidal effect of this wavelength has so far discouraged attempts at absorption imaging of chromatin dynamics in living cells. However, estimates based on our results indicate that absorption imaging can achieve contrasts up to $\approx$30\% in 100 nm chromatin fibres and $\approx$3\% contrast for 30 nm chromatin fibres, sufficient to detect real time dynamics over tens of image frames with sub-lethal dosing. The time resolution of absorption imaging is limited by the signal to noise ratio required to recover the object contrast from the photon shot noise in each pixel. The maximum frame rate is limited by available camera technology to kilohertz scales. Molecular optical transitions have higher saturation intensities compared with the strong atomic transition used in this work, indicating ms timescale resolution would  be feasible with a suitable illumination pulse sequence.\\

\textbf{Methods}\\
\begin{small}
\textbf{An ideal scatterer}\\
For an ideal two level atom the scattering rate is
\begin{equation}
\gamma_p= \frac{\Gamma}{2}\frac{ s_0 }{ 1+ s_0 + 4\delta^2/\Gamma^2 }
\end{equation}
where $\Gamma$ is the natural linewidth of the transition, $s_0=I/I_{\mbox{sat}}$ is the normalised saturation intensity, and  $\delta$ is the laser detuning. The power scattered ($P_{scat}$) by the absorption of photons in such an atom is $P_{scat}= \gamma_p h c /\lambda$. The power scattered approaches $P_{\mbox{max}}$ in the limit of high intensity. The 370 nm transition in $^{174}$Yb$^+$ closely follows the predictions of a two-level semi-classical model due to its lack of hyperfine structure (I=0) and unresolved Zeeman structure. $^{174}$Yb$^+$ ions were produced and laser cooled in a double-needle RF trap using an all diode laser setup. The apparatus has been previously documented in \cite{Kielpinski-06, Streed-08, Streed-11, Jechow-11}. For each photon scattering event, there is a small probability (0.5\%) that the ion decays into the meta-stable ($\tau$=53ms) dark $D_{3/2}$ state \cite{Yu-00}. To maintain a high scattering rate the ions were repumped back to the S$_{1/2}$ state by driving the transition at 935.2 nm. Deliberate interruption of the 935.2 nm laser beam pumped the ion into the dark $D_{3/2}$ state and was used to reduce the scattering of 369.5 nm light by the ion by an estimated 45 dB. Illumination of the ion for absorption imaging resulted in charging of the in-vacuum dielectric lens surfaces near the ion, similar to those previously reported for anti-reflection coated glass \cite{Harlander-10}. The resulting DC electric fields displace the ion from the trap centre and were compensated by applying voltages on four additional needles located close to the trapping region. Our system's high magnification is particularly useful in measuring the ion's displacement from the trap centre. Low illumination beam powers of 15 nW ($570  \: \mbox{W} \: \mbox{m}^{-2}$) resulted in a slow charging rate whose effect was noticeable after several hours. Above 100 nW the charging rate became problematic with noticeable degradation occurring within minutes. Powers greater than 130 nW rapidly resulted in a loss of the ion.

\textbf{Image Analysis}
\begin{figure}[htbp]
\begin{center}
\includegraphics[width=120mm]{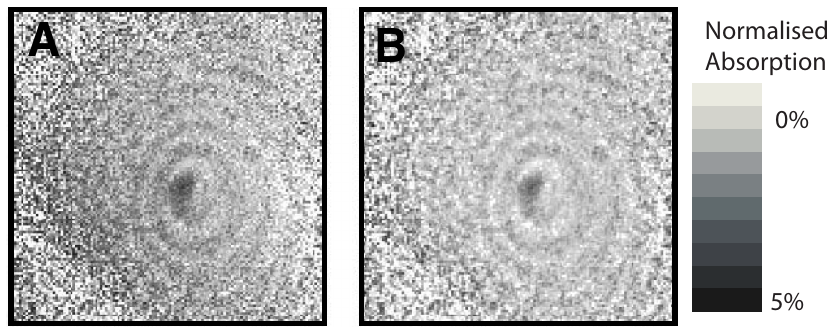}
\caption{Absorption images before (A) and after (B) filtering procedure. The illumination spot intensity is greater than the camera read noise across most of the chip.}
\end{center}
\end{figure}

Signal and background images were acquired using the Andor MCD software package and were analysed in Mathematica\textcopyright. To improve the success rate for the two dimensional fitting routine in the lower signal data (high intensity or far detuning), the images were bandpass filtered to reduce noise using a two step process. Figure 4 provides images of the absorption profile before (A) and after (B) filtering. High spatial frequency noise such as hot pixels and read noise were first removed with the Mathematica\textcopyright$\:$ GaussianFilter image processing function with a 1 pixel radius. Low spatial frequency components were then removed by subtracting a GaussianFilter smoothed image with 20 pixel radius (approximately double the 10 pixel feature size) from the original image. These values were determined empirically such that the contrast values for high contrast images ($>2$\%) were unaffected and stable fits could be obtained for the lower contrast images. The consistency of the fit was verified by checking the agreement in ion image widths and positions across the complete data set.

\newpage

\noindent \textbf{Acknowledgements}
\noindent Supported by the Australian Research Council under DP0773354 (DK), DP0877936 (EWS, Australian Postdoctoral Fellowship), and FF0458313 (H. Wiseman, Federation Fellowship) as well as the US Air Force Office of Scientific Research (FA2386-09-1-4015). AJ is supported by a Griffith University Postdoctoral Fellowship. The phase Fresnel lens was fabricated by M. Ferstl at the Heinrich-Hertz-Institut of the Fraunhofer-Institut f\"{u}r Nachrichtentechnik in Germany.\\

\noindent \textbf{Author contributions}
\noindent EWS and DK designed the experiment. EWS, AJ, BGN, and DK constructed the apparatus. EWS, AJ, BGN, and DK conducted the experiment and collected the data. EWS, AJ, and BGN analysed the data. The manuscript was prepared by EWS with contributions from AJ, BGN, and DK.\\

\noindent \textbf{Additional information}
\noindent The authors declare that they have no competing financial interests. Reprints and permissions information is available online at http://npg.nature.com/reprintsandpermissions/. Correspondence and requests for materials should be addressed to DK~(email: d.kielpinski@griffith.edu.au).
\end{small} \\

% must be less than 350 words
\noindent \textbf{Figure legends} \\
\noindent \textbf{Figure 1. Configuration of experimental apparatus.} A laser cooled $^{174}$Yb$^+$ ion is confined in a radio frequency Paul trap formed by the electric quadrupole (dashed lines) between two tungsten needles. Resonant illumination at $\lambda$=369.5nm is focused to a spot 4.8 $\mu$m FWHM and absorbed by the ion. The resulting shadow is imaged with a large aperture phase Fresnel objective lens onto a cooled CCD camera at $615\times$ magnification (additional optics omitted for clarity).\\
% 95 words
\noindent \textbf{Figure 2. Comparison between absorption and fluorescence images.} Absorption (A,B) and fluorescence (C,D) image data of a single isolated $^{174}$Yb$^+$ ion. Arrows in the absorption image (B) and fluorescence image (D) indicate the position of horizontal cross-sections (A) and (C). Measured at -8 MHz from resonance, near the optimum Doppler cooling detuning for the 369.5 nm transition. Fits are to a two dimensional gaussian over the the whole image. Exposure times are 1 s for B (4x4 pixel binning) and 60 s for D (no binning). Images (B,D) are 3.1 $\mu$m across.\\
\noindent \textbf{Figure 3. Dependence of absorption properties on laser parameters.} The contrast was measured as a function of laser detuning (A) and incident intensity (B) with the other parameter fixed near its optimum value (dashed vertical lines at -8 MHz and $570 \: \mbox{W} \: \mbox{m}^{-2}$ on the alternate graphs). At 0 MHz detuning (rightmost point in (A)) laser cooling ceased and the ion was not detectable until recooled. (C) Total power absorbed by the ion as a function of the incident power on the maximum scattering cross section $\sigma_0$. Dashed horizontal line is the theoretical maximum power absorption $P_{\mbox{max}}= hc/(2\lambda\tau)$. Solid line is the linear extrapolation of the fit at low intensity. Each data point is obtained by fitting to a single image with error bars from the statistical uncertainty in the fit.\\
\noindent \textbf{Figure 4. Spatial filtering of absorption images.} Absorption images before (A) and after (B) filtering procedure. The illumination spot intensity is greater than the camera read noise across most of the chip. Images are 10.6 $\mu$m across.\\

\end{document}